\documentclass[lettersize,journal]{IEEEtran}
\usepackage{amsmath,amsfonts}
\usepackage[x11names]{xcolor}
\usepackage{algorithmic}
\usepackage{array}
\usepackage[caption=false,font=normalsize,labelfont=sf,textfont=sf]{subfig}
\usepackage{textcomp}
\usepackage{stfloats}
\usepackage{url}
\usepackage{verbatim}
\usepackage{graphicx}
\usepackage{hyperref}
\hypersetup{
    colorlinks=false,
    }
\def\BibTeX{{\rm B\kern-.05em{\sc i\kern-.025em b}\kern-.08em
    T\kern-.1667em\lower.7ex\hbox{E}\kern-.125emX}}
\usepackage{balance}
\usepackage{cite}
\begin{document}

\newcommand{\yun}[1]{\textcolor{blue}{YUN: #1}}

\title{Automatic Retrieval-augmented Generation of 6G Network Specifications for Use Cases}
\author{Yun Tang, Weisi Guo
\thanks{Authors are with Cranfield University, UK. This work is supported by EPSRC CHEDDAR: Communications Hub For Empowering Distributed ClouD Computing Applications And Research (EP/X040518/1) (EP/Y037421/1). 
All community-contributed use cases and specifications are openly available at our website: \url{https://ntutangyun.github.io/llm_6g_frontend}.
}
}


\maketitle


\begin{abstract}
6G Open Radio Access Networks (O-RAN) promises to open data interfaces to enable plug-and-play service Apps, many of which are consumer and business-facing. Opening up 6G access lowers the barrier to innovation but raises the challenge that the required communication specifications are not fully known to all service designers. As such, business innovators must either be familiar with 6G standards or consult with experts. Enabling consistent, unbiased, rapid, and low-cost requirement assessment and specification generation is crucial to the O-RAN innovation ecosystem.

Here, we discuss our initiative to bridge service specification gaps between network service providers and business innovators leveraging Large Language Models (LLMs). We first review the state-of-the-art and motivation in 6G plug-and-play services, capabilities, potential use cases and LLMs. We identify an ample innovation space for hybrid use cases that may require diverse and variational wireless functionalities across its operating time. We show that the network specification can be automated and present the first automatic retrieval-augmented network service specification framework for 6G use cases. To enable public acceptance and feedback, a website interface is published for the research and industrial community to experiment with the framework. We hope this review highlights the need for emerging foundation models for this area and motivates researcher engagement and contribution to the community through our framework.

\end{abstract}


\section{Introduction}

\IEEEPARstart{T}{he} future generation (6G) of communication networks are envisioned to push the boundary of networking and computation capabilities, supporting cutting-edge services such as Hyper-reliable Low-Latency Communication, Massive Communication and Ubiquitous Connectivity \cite{IMT_2030, NGMN_6g_use_cases}. One major proposed area is the opening up of data interfaces in the Radio Access Network. 
The O-RAN ecosystem enables a wide range of internal RAN management and external B2B services for diverse use cases, encouraging innovation from start-ups and smaller enterprises \cite{ORAN, BubbleRAN}.

The open challenge is that some innovators may not understand the detailed 6G network specifications available based on their business needs. Each use case may have several sub-tasks that require different and evolving 6G network specifications. Traditionally, the designer of these use cases would need to seek in-house or external consultancy with telecommunication experts (e.g., Ericsson). However, this can prove costly, incur delays, and may result in variational advice, deterring many innovators.

\begin{figure}[t]
    \centering
    \includegraphics[width=\columnwidth]{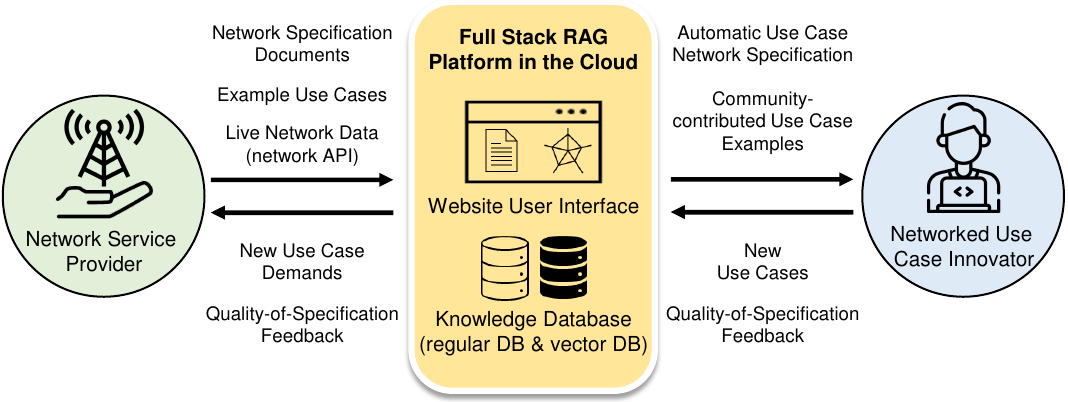}
    \caption{Functional overview of the full-stack platform with a website frontend and knowledge database backend bridging and assisting network service providers and networked use case innovators.}
    \label{fig:platform_overview}
\end{figure}


On the other hand, the network service providers must thoroughly consider the expected use case demands when planning the optimal allocation of network infrastructures (edge nodes and base stations) and resources (bandwidth and data rate). While existing use cases are categorized \cite{NGMN_6g_use_cases} based on the predominant network specifications (e.g., \textit{telemedicine} is often placed under \textit{hyper-reliable low-latency communication} category), the emergence of hybrid or rare use cases with mixed or dynamic network requirements poses open challenges.

From the use case innovator's point of view, an automated process of identifying the appropriate balance of network specifications based on the use case description and the current accessible networking capabilities would aid in the innovation and development of user applications. Meanwhile, from the network service provider's point of view, a collection of concrete use cases with comprehensive network service specifications would also significantly benefit the implementation of resource management plans.

Addressing the demand above, this article presents our initiative to construct a knowledge database with a public interface for the design of future diverse and dynamic networked use cases, leveraging the power of retrieval-augmented Large Language Modes (LLMs) as shown in Fig.~\ref{fig:platform_overview}. Specifically, we first review the potential 6G use cases, corresponding network service specifications, and the limitations of the current specification descriptions (Section~\ref{sec:use_case_review}). Then, we review the state-of-the-art (SOTA) LLMs and their applications in the 6G network (Section~\ref{sec:llm_review}). Informed by the literature review, we design a crowd-sourced knowledge database of use cases with their specifications that powers a RAG framework to enable automatic use case extraction and specification generation (Section~\ref{sec:methodology}).
\section{6G Capabilities, Services, and Use Cases}\label{sec:use_case_review}

\subsection{6G Network Capabilities}

\begin{figure}
    \centering
    \includegraphics[width=\columnwidth]{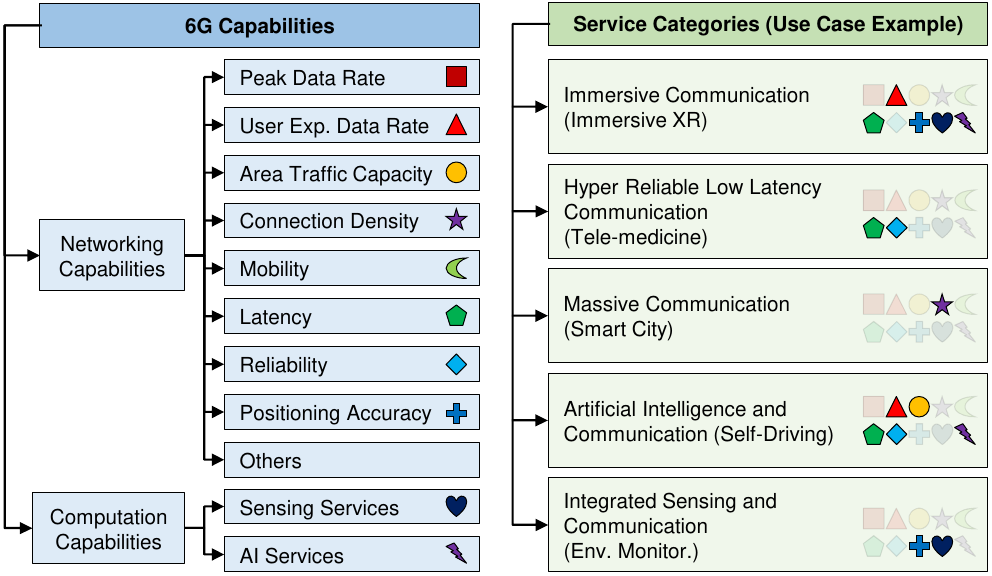}
    \caption{6G network capabilities and service categories with a use case example and primary network capability specifications highlighted. Adapted from \cite{IMT_2030}.}
    \label{fig:6g_capabilities_and_use_cases}
\end{figure}

As discussed in the IMT 2030 standard \cite{IMT_2030} (Fig.~\ref{fig:6g_capabilities_and_use_cases}) and many other literature, 6G is envisioned to excel in many networking capabilities compared to precedent generations, specifically:
\begin{itemize}
    \item Peak data rate: the highest data rate achievable per device can be up to 200 Gbit/s.
    \item User experienced data rate: ubiquitously available data rate per device can be up to 500Mbit/s.
    \item Area traffic capacity: total traffic throughput per geographic area can be up to 50 Mbit/s/m$^2$.
    \item Connection density: total number of connected devices per unit area can be up to $10^8$ devices/km$^2$.
    \item Mobility: the maximum speed with defined QoS can be up to 1000 km/h.
    \item Latency: the minimum latency over the air interface can be as small as 0.1 ms.
    \item Reliability: the probability for successfully transmitting a predefined amount of data within a predetermined time duration can be as high as $1-10^{-7}$.
    \item Positioning accuracy: the approximated position of connected devices can be as accurate as 1 cm.
    \item Other: details of other non-user-facing capabilities such as spectrum efficiency can be found in the standard \cite{IMT_2030}.
\end{itemize}

In addition to the enhanced networking capabilities, we will in future (see Section~\ref{sec:future}) also include 6G capabilities in AI, privacy, security, and sensing.

\subsection{B2B 6G Services and Use Cases} 

The major driving forces pushing the capability boundaries are the use cases. Different use cases demand different sets of networking capabilities. Although it might be uncommon, if not infeasible, for a single use case to have the highest requirement in every specification facet, it is pretty common to have a group (i.e., a service category) of use cases sharing the same specification pattern, as shown in Fig.~\ref{fig:6g_capabilities_and_use_cases}, where the service categories are often named by the predominant capability specifications. For example, the \textit{Immersive Communication} category contains use cases such as immersive XR and holographic telepresence requiring the transmission of high-resolution video, audio, holograms or haptic-type data and accurate position/gesture estimations at real-time latency. As a result, the use cases of this category share the same specification pattern of high data rate, high positioning accuracy, and real-time low latency. In contrast, the use cases of the \textit{Massive Communication} category highly depend on the connection density with relatively softer latency and reliability constraints.

Such a high-level categorization approach helps identify the most stringent network service requirements. However, mapping a use case with a single requirement pattern, as shown in Fig.~\ref{fig:6g_capabilities_and_use_cases}, would hinder optimal networking and computation resource allocation due to the limitations below. 

\textbf{Limitation 1: Disregard for Functional Diversity} New use cases may arise which do not belong to any specific category but, instead, have hybrid specification patterns as a single use case often initiates multiple wireless connections for different functionalities. For example, an autonomous vehicle may have one \textit{integrated sensing and communication} connection with edge nodes for collaborative sensing, one \textit{hyper-reliable low-latency communication} connection with surrounding traffic participants for collaborative motion planning and another regular connection for occasional system monitoring. It would be a waste of resources if the monitoring function shared the same hyper-reliable, low-latency connection channel.

\textbf{Limitation 2: Disregard for Temporal Diversity} networking specifications may vary over time as new functional connections arise during the use case life-cycle. Take the autonomous driving use case, for example. An ad hoc \textit{immersive communication} connection with a remote fallback operator may be established upon encountering emergencies with entirely different networking specifications from the existing connections above. Such temporal diversity requires dynamic reallocation of networking resources in real-time for optimal QoS and economy and hence also demands due consideration.

\subsection{Native Services in O-RAN} 

It is expected that many of the native services will be implemented as xApps or rApps in the O-RAN, with the primary goal of optimizing the networking resources based on the real-time use case specifications. xApps are installed on near real-time RAN intelligent controllers (Near-RT RIC) and perform sub-second network optimization tasks such as connection management \cite{orhan2021connection}. On the other hand, rApps are installed on non-real-time RIC (Non-RT RIC) and focus on broader network management strategy and longer-term network planning \cite{qazzaz2024machine}. Regardless of the implemented optimization algorithms, whether pre-fixed control policies from look-up tables or ML/AI techniques, we identify the following limitation during the current O-RAN app development process.


\textbf{Limitation 3 Limited sharing of design references} Current O-RAN apps are often designed based on a limited set of manually crafted hypothetical use cases, which cannot reflect the diverse demands of the use case innovation community. Such a lack of up-to-date references hinders the design and deployment process and thus leads to poor performance for early adopters.
\section{LLMs and 6G}\label{sec:llm_review}



\subsection{Introduction to LLMs}
LLMs are language models for natural language processing (understanding natural language input and generating natural language output). Modelled by billions of parameters and trained on vast corpora of text data, LLMs can learn intricate patterns of language usage and knowledge across disciplines. As a result, they have shown remarkable performance across a wide range of tasks, including machine translation, text summarization, sentiment analysis, and even more complex applications like conversational AI and code generation, and thus garnered widespread adoption across numerous application domains. Worth noting is that they can convert unstructured natural language into formal specifications \cite{LLM_logic}. Through fine-tuning and a human-in-the-loop, it can pave a pathway towards being compliant with IEEE property specification language (PSL), Signal Temporal Logic (STL), or System Verilog Assertions.

New LLMs are developed and released almost weekly and the SOTA LLMs are GPT-4o, Llama 3.1 405B, Claude 3.5 Sonnet, Gemini 1.5 Pro, Llama 3.1 (70B) and Mistral Large 2, demonstrating exceptional capabilities across various benchmarks on general language tasks\footnote{See example comparison of AI models across quality, performance and price: artificialanalysis.ai}. 

Even the best LLMs make mistakes due to inevitable misinformation in the training data. In addition, LLMs are known to hallucinate, generating plausible but incorrect or nonsensical information, which can be particularly problematic when they confidently present fabricated facts or details that have no basis in reality. Leveraging the in-context learning capability, the input prompt to LLMs can be augmented with validated references (i.e., retrieval-augmented generation) which can significantly mitigate the hallucination issue \cite{lewis2020retrieval}.

\subsection{LLMs and 6G}

Researchers have made efforts to harness the power of LLMs in the era of 6G networks focusing on three main application topics:

\textbf{Topic 1: 6G Network Design for LLMs} where the 6G network architecture is explicitly customised for embedding LLMs. For example, \cite{lin2023pushing} envisioned a 6G mobile edge computing (MEC) architecture consisting of \textit{network management component} for thering global network knowledge and orchestrating model training/inference across the edge nodes and \textit{edge model caching component} for caching LLM models at edge nodes, and discussed techniques such as distributed learning and split inference to enable training and inference at RAN nodes with limited computation resources.

\textbf{Topic 2: LLM for B2B 6G Services} where the 6G network empowers LLMs to offer intelligence services to 6G customer's use cases. For example, \cite{LLM_6G} proposes a split learning framework where small LLMs ($\leq$10B) are installed through RAN connection on end user's mobile devices to provide multi-modal user input interface and execute the intelligence tasks allocated by large LLMs ($>$10B) installed on edge (or cloud) servers. The large LLMs maintain a Digital Twin (DT) connected to an internal and external knowledge database for decision grounding. In general, this application topic mainly focuses on the following areas \cite{LLM_6G}:
\begin{enumerate}
    \item How do distributed LLM services across the O-RAN enable user access?
    \item How can LLMs enable reliable digital twins (DTs) that capture reasoning, planning, verification and reflection?
    \item How to create context-aware (e.g., semantic, task-aware) communication or integrated sensing?
\end{enumerate}

\textbf{Topic 3: LLM for Native 6G Services} where LLMs are utilised (e.g., in xApps and rApps) for autonomous 6G network management. The general workflow is that: first, the LLMs perceive the communication environment semantics by parsing the multi-modal input data either from user input or network sensors; then, LLMs make informed decisions in real-time to adjust networking parameters or allocate networking resources to improve communication QoS; and last, LLMs keep evolving themselves by distributed learning techniques from past experiences and new scenarios. 

In fact, works under Topic 3 can be regarded as indirect enablers for those in Topic 2 since, after all, the 6G network capabilities are enhanced for and evaluated by the QoS of serviced use cases. Hence, the focal point is the definition of networked use cases and their dynamic networking specifications, and we hereby identify the following limitation: 

\textbf{Limitation 4 Absence of Automated Network Specification Tool} In all these cases, the general assumption is that 6G already has a set of services (semantic-aware ISAC, DTs, edge LLM access) and well-defined specifications (data rate, latency, reliability), and the use cases engage with these services and capabilities in well-defined ways. We expect 6G O-RAN to attract many innovators with \textit{plug-and-play} business use cases of a diverse range of demands \cite{ORAN}. However, what if the use case innovators do not know what services or capabilities they need or are available in the designing or operating phase? Reading the network specification documents does not solve this due to the following limitations: 1) the documents do not cover every diverse use case; 2) the existing documents are often network service provider-oriented discussing the network architecture designs instead of use case innovator-focused discussing concrete use case networking requirements, and 3) the documents cannot offer live resource information available in the network. In these situations, we may need LLMs to play a different role, one where they help the use case innovators to specify 6G network requirements in real time based on the usage context and up-to-date network knowledge.

\section{Prototype Operation}\label{sec:methodology}

\begin{figure*}
    \centering
    \includegraphics[width=\textwidth]{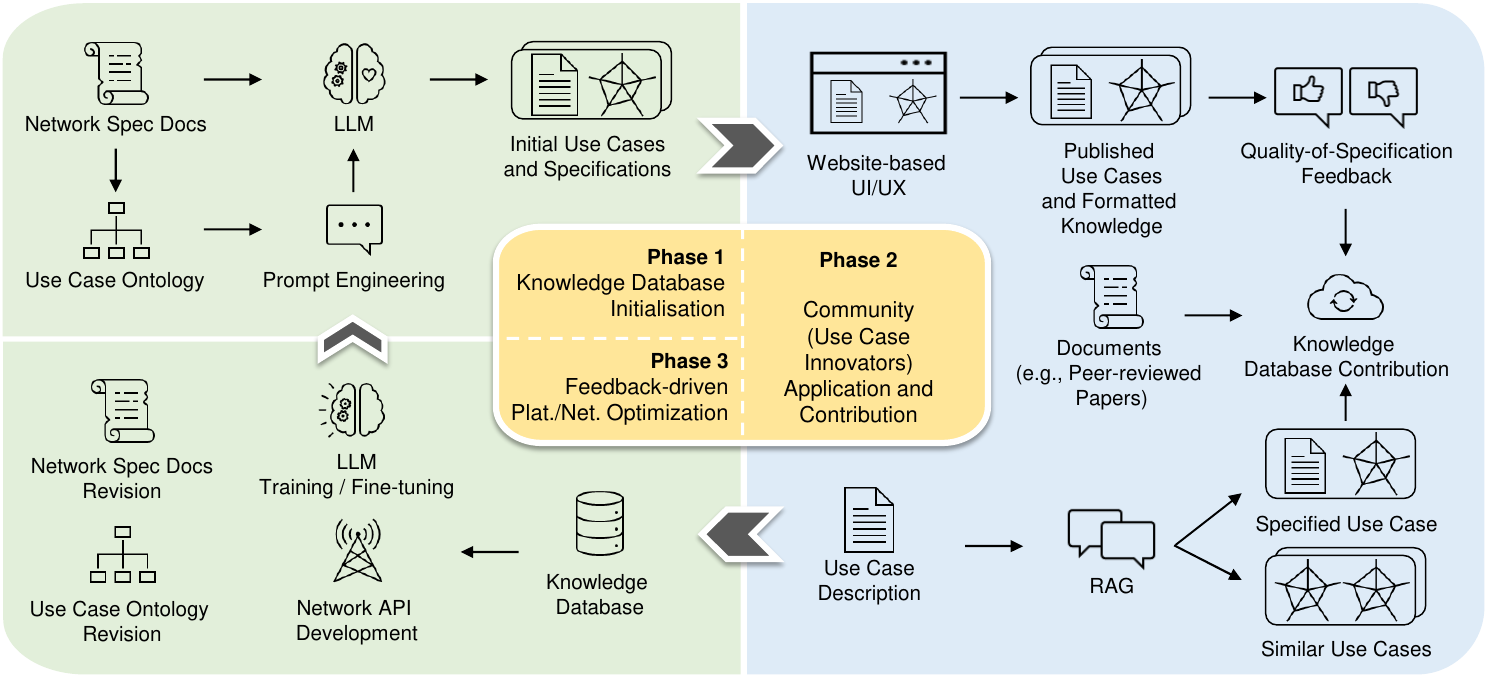}
    \caption{Overview of platform operation phases (green for network service providers and blue for use case innovators): (Phase 1) network service provider initializes the knowledge database; (Phase 2) the community benefits from the automated specification generation and contributes with new use cases and feedbacks; and (Phase 3) the network service operator benefits from the community contributions for platform update and network development.}
    \label{fig:operation_phases}
\end{figure*}

To address the limitations above, we initiate a crowd-sourced knowledge database for 6G use cases and implement an extensible user interface for engaging the stakeholders. Fig.~\ref{fig:operation_phases} details the three operation phases of the platform for both network service providers (Phase 1 and 3) and use case innovators (Phase 2), which are discussed in a top-down fashion in the following subsections. The RAG pipeline is built with the open-source LlamaIndex library and the knowledge database is powered by MongoDB Atlas Database.

\subsection{Phase 1 Knowledge Database Initialisation}
This phase constructs the knowledge database with seed use cases and corresponding network specifications. LLMs are utilised via prompt engineering to extract the use case knowledge from the given documents (network standards, white papers, publications, etc.) and format them according to the use case ontology.

\begin{figure}
    \centering
    \includegraphics{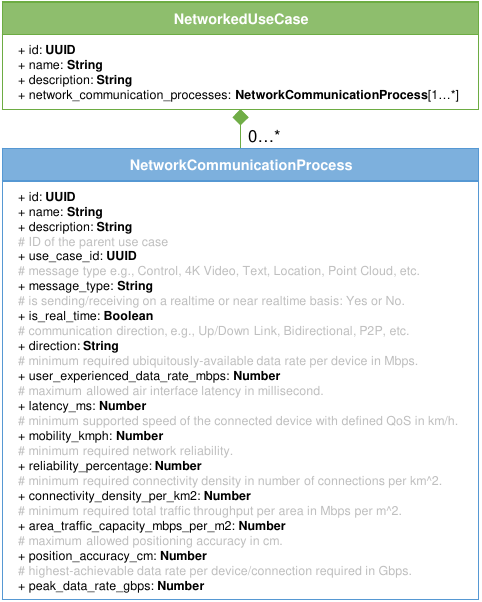}
    \caption{Use case ontology presented as UML class diagrams.}
    \label{fig:use-case-ontology}
\end{figure}

\textbf{Use Case Ontology} As discussed in Limitations 1 and 2, a single use case comprises a dynamic list of networked communication sub-processes, each requiring distinct network specifications tailored to its individual needs. To embrace such flexibility, we design a minimalist ontology presented in Fig.~\ref{fig:use-case-ontology} for use cases and their networked communication processes, modelled in Unified Modeling Language (UML) for easy implementation and extension.
The ontology can be customized manually by networking experts or automatically using LLMs \cite{tang2023domain}. 
Both use cases and their communication processes are indexed by the \textit{id} attribute and described by the \textit{name} and \textit{description} attributes, which are essential for semantic matching in the RAG retrieval phase. A use case contains multiple communication processes. Each communication process represents a single wireless connection served by a single network slice for a single communication purpose established at any time (\textit{is\_real\_time}) during the use case life-cycle for either transmitting or receiving (\textit{direction}) messages of a specific type (\textit{message\_type}). These three metadata attributes (\textit{is\_real\_time}, \textit{direction} and \textit{message\_type}) are explicitly introduced to assist the inference of the remaining network specification attribute values leveraging the chain-of-thought philosophy. Eight typical network specification metrics for wireless use cases are considered, i.e., \textit{user\_experienced\_data\_rate\_mbps}, \textit{latency\_ms}, \textit{mobility\_kmps}, \textit{reliability\_percentage}, \textit{connectivity\_density\_per\_m2}, \textit{area\_traffic\_capacity\_mbps\_per\_m2}, \textit{position\_accuracy\_cm} and \textit{peak\_data\_rate\_gbps}, whose value ranges can be configured by network service providers or automatically derived from the initialisation documents utilising LLMs.

\textbf{Knowledge Database Design} The knowledge database comprises a regular and a vector database, storing the raw textual data and their corresponding vector embedding, respectively. Hence, the vector database is used for semantic matching during the RAG retrieval process, and the raw textual data of the matched use cases are then extracted from the regular database to compose the final prompt context.

\textbf{UI/UX Design} A website-based user interface aims to facilitate the use case ontology configuration, document submission, LLM selection, prompt customisation and use case validation in the initialisation phase.

\subsection{Phase 2 Global Innovator Application and Contribution}

Through the website interface, use case innovators can 1) view and comment (or vote) the published use cases and extracted knowledge; 2) share their peer-reviewed papers for knowledge extraction; 3) query the network specification through RAG with their use case description; and 4) contribute their use cases. The user-contributed use cases and documents aim to be automatically verified by LLMs or manually by domain experts before being merged into the knowledge database, to ensure the authenticity of the knowledge database, and based on this, the trustworthiness of the generated specification.

\begin{figure*}
    \centering
    \includegraphics[width=\textwidth]{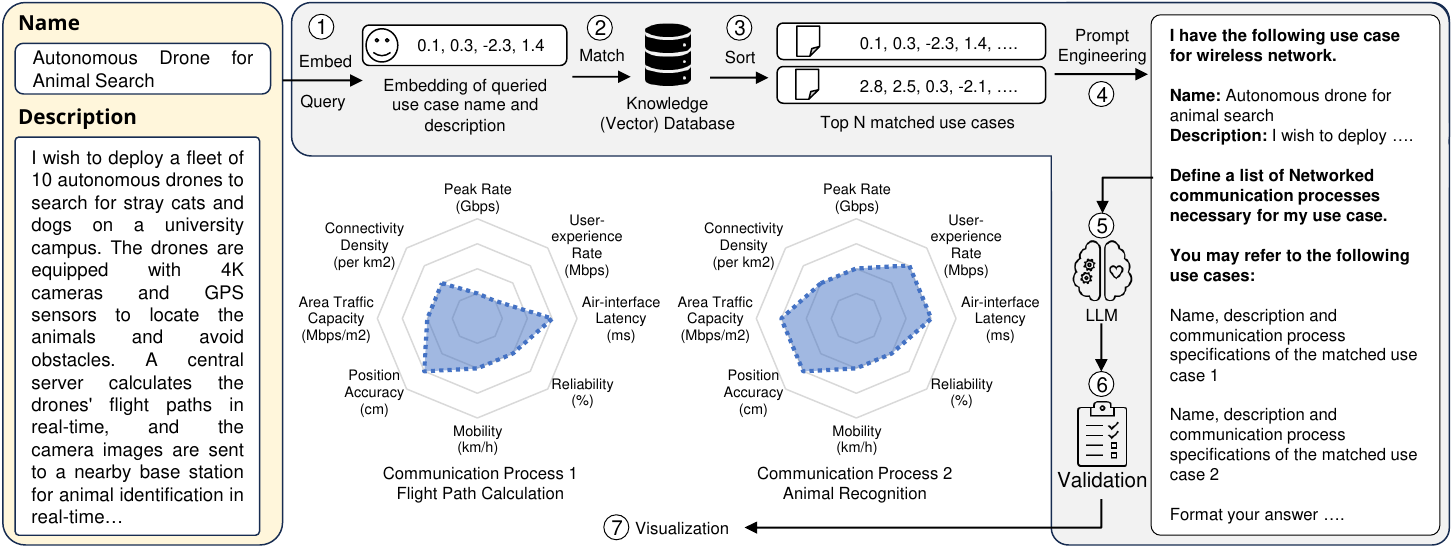}
    \caption{Illustration of the core RAG workflow for automatic network specification generation for user-described use cases.}
    \label{fig:automatic-spec-gen-workflow}
\end{figure*}

\textbf{Automatic Use Case Specification Generation} For novice use case developers, a concise description of the intended networked application is all the system needs to suggest the necessary communication processes and infer the corresponding network specifications. In addition, similar use cases matched during the RAG retrieval process will also be returned to the developer for reference.
Fig.~\ref{fig:automatic-spec-gen-workflow} illustrates the core workflow of the RAG process during automatic network specification generation including the following steps: 

\begin{enumerate}
    \item The name and description of the use case are embedded into numeric vectors via embedding models (e.g., \textit{text-embedding-ada-002} model by OpenAI). The vector captures the semantic meaning of the queried use case description.
    
    \item All the use cases stored in the knowledge database are matched with the query and sorted by the semantic distance (e.g., via cosine similarity). Due to the limited context length, only the top N matched use cases from the knowledge database are selected as the final retrieval results.
    
    \item Appropriate prompt engineering (e.g., templates) can be applied to integrate the queried and retrieved use cases into a single LLM prompt and feed into an LLM for RAG-informed generation.

    \item The use case ontology is used to derive the desired LLM output format and implement the output parser, which returns a list of structured communication processes. 

    \item Radar (or spider) charts are used to visualise each communication process's numerical network specification metrics.
\end{enumerate}

Advanced prompt engineering techniques such as user query pre-processing and retrieval result post-process (e.g., re-ranking, transformation, filtering) are applicable but omitted for brevity in the illustration.

\subsection{Phase 3 Feedback-driven Platform and Network Optimization}

The knowledge database aims to bridge the communication gap between network service providers and use case innovators in both ways. The community-contributed use cases and feedback are valuable references for process updates of the platform in the short run and the network in the long run. For example, network service providers can fix errors or reorganize the network specification documents and use case ontology based on the retrieval statistics and user comments on the extracted knowledge; LLMs can be trained or fine-tuned through reinforcement learning with human feedback. The network service provider may also develop on-demand northbound APIs (as specified in 3GPP CAPIF Framework, TMF IF 1167, and other publications of 5G-PPP Software Network WG and The Linux Foundation Telco Global API Appliance Project) for sharing the network's resource availability or allocation data \cite{charismiadis20233gpp} for better generation of use case specifications. In the long run, the collected use cases and specifications offer first-hand insights into network infrastructure provisioning and network resource allocation plans.

\section{Community Engagement \& Future}\label{sec:future}

\begin{figure}
    \centering
    {\setlength{\fboxsep}{2pt}\setlength{\fboxrule}{0.1pt}
    \fbox{\includegraphics[width=0.97\columnwidth]{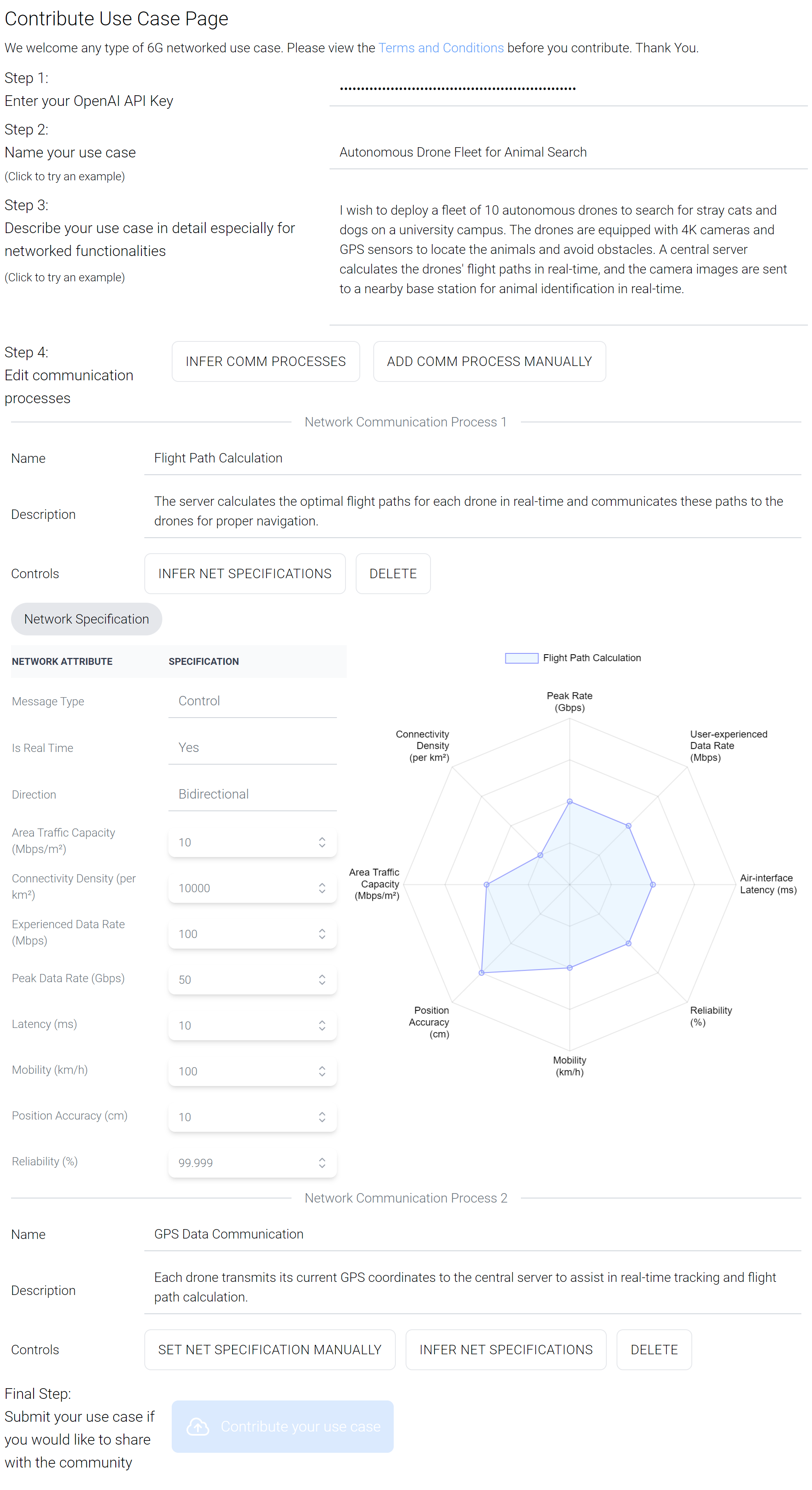}}}
    \caption{Screenshot of our website (\url{https://ntutangyun.github.io/llm_6g_frontend}) allowing the website user to perform RAG-informed automatic network specification generation and visualization for the described use case and contribute the use case to the public knowledge database.}
    \label{fig:website-screenshot}
\end{figure}

Different from existing tools such as Nvidia O-RAN ChatBot \cite{nvidia_oran_chatbot_multimodal} or Telco-RAG \cite{bornea2024telco} which focus on answering factual questions of the telecom documents, our goal is to showcase RAG-based automation for future O-RAN 6G service innovation. Currently, website users can perform automatic specification retrieval-augmented generation of their described use cases utilizing the knowledge database. We expect the full-stack website, along with the knowledge database, to be rolled out in the following four stages of functionality:
\begin{enumerate}
    \item Stage I - User Trials (public): Fig.~\ref{fig:website-screenshot} shows the screenshots of our tool for community trial. One may check out the existing RAG knowledge database, including all the current use cases, their associated telecommunication processes and specifications. One may also contribute their own use cases, which will undergo an automatic RAG-based evaluation and manual confirmation before merging into the knowledge database. Currently, it requires an OpenAI API account, and in the future, we may develop and publish our own locally trained or fine-tuned LLM as we gather a sufficient amount of knowledge in the database to reduce the cost barrier for non-commercial users.
    \item Stage II - Crowd-sourcing Knowledge Contributions (available but not published yet): Fig.~\ref{fig:operation_phases} shows the crowd-sourcing element, whereby expert users can upload technical papers, specifications, and evidence of how different use cases map to 6G functionalities and specifications. This would enrich the RAG and provide more fine-scale tuning of the LLM performance. Currently, this functionality is available but not visible on the website because we need to develop appropriate mechanisms to moderate and ensure high-quality content. 
    \item Stage III - Tertiary Requirements: we will add tertiary requirements to the system related to cyber security and AI. Some of these have formal specifications and requirements, and some do not, and the challenge is to design appropriate RAG databases to reflect this.
    \item Stage IV - Comprehensive Evaluation: current quantitative metrics such as hallucination or BLEU scores only evaluate the RAG performance of (different) LLMs in terms of retrieval-wise accuracy according to the indexed documents in the knowledge database. However, the retrieval-wise accuracy of RAG on the generated specifications (e.g., latency, reliability or security) does not necessarily mean optimal performance after deployment as the indexed knowledge can be obsolete or sub-optimal. We will extend the feedback loop to use case deployment and explore evaluation methodologies for operation-wise accuracy. In addition, we also plan to conduct extensive studies to evaluate the time-saving and cost-saving benefits compared to specification documents reading and telecom expert consulting.
\end{enumerate}

We call for community feedback with votes and comments, as well as the contribution of novel use case specifications and documentation to enrich the shared knowledge database in subsequent phases.

Looking into the future, we hope the research can inform real-time network control capabilities where dynamic service specifications can be generated faster and drive semantic-based network optimisation \cite{LLM_control}. For example, the LLM outputs may inform/contextualise the attention space of reinforcement-based RAN Intelligent Controller (RIC).

\section{Conclusion}

6G future networks are envisioned to surpass previous generations in many networking capabilities and provide ubiquitous computation resources for the network and the networked use cases. However, there is a knowledge gap between the network service providers and the use case innovators regarding the detailed network service specifications for the ever-changing use cases. To bridge the gap, we propose a public future use case knowledge database initiative and implement a RAG-empowered automatic network specification generation prototype for the telecommunication community. We have shown not only the relevant literature around the technologies in this area (ORAN Apps, 6G, RAG-based LLMs) but also designed and publicly released a prototype.

We hope the framework will prove beneficial to network service providers in infrastructure planning and resource allocation, use case innovators in the development of future use cases, and the entire telecommunication community towards knowledge-empowered autonomous future networks.

\bibliographystyle{IEEEtran}
\bibliography{Main}

\begin{thebibliography}{10}
\providecommand{\url}[1]{#1}
\csname url@samestyle\endcsname
\providecommand{\newblock}{\relax}
\providecommand{\bibinfo}[2]{#2}
\providecommand{\BIBentrySTDinterwordspacing}{\spaceskip=0pt\relax}
\providecommand{\BIBentryALTinterwordstretchfactor}{4}
\providecommand{\BIBentryALTinterwordspacing}{\spaceskip=\fontdimen2\font plus
\BIBentryALTinterwordstretchfactor\fontdimen3\font minus \fontdimen4\font\relax}
\providecommand{\BIBforeignlanguage}[2]{{%
\expandafter\ifx\csname l@#1\endcsname\relax
\typeout{** WARNING: IEEEtran.bst: No hyphenation pattern has been}%
\typeout{** loaded for the language `#1'. Using the pattern for}%
\typeout{** the default language instead.}%
\else
\language=\csname l@#1\endcsname
\fi
#2}}
\providecommand{\BIBdecl}{\relax}
\BIBdecl

\bibitem{IMT_2030}
\BIBentryALTinterwordspacing
ITU, ``{M.2160: Framework and overall objectives of the future development of IMT for 2030 and beyond},'' 2023, accessed on 14.08.2024. [Online]. Available: \url{https://www.itu.int/rec/R-REC-M.2160-0-202311-I/en}
\BIBentrySTDinterwordspacing

\bibitem{NGMN_6g_use_cases}
\BIBentryALTinterwordspacing
NGMN, ``{6G use cases and analysis},'' 2023, accessed on 14.08.2024. [Online]. Available: \url{https://www.ngmn.org/publications/6g-use-cases-and-analysis.html}
\BIBentrySTDinterwordspacing

\bibitem{ORAN}
M.~Polese, M.~Dohler, F.~Dressler, M.~Erol-Kantarci, R.~Jana, R.~Knopp, and T.~Melodia, ``Empowering the 6g cellular architecture with open ran,'' \emph{IEEE Journal on Selected Areas in Communications}, vol.~42, no.~2, pp. 245--262, 2024.

\bibitem{BubbleRAN}
F.~A. Bimo, R.-G. Cheng, C.-C. Tseng, C.-R. Chiang, C.-H. Huang, and X.-W. Lin, ``Design and implementation of next-generation research platforms,'' in \emph{IEEE Globecom}, 2023.

\bibitem{orhan2021connection}
O.~Orhan, V.~N. Swamy, T.~Tetzlaff, M.~Nassar, H.~Nikopour, and S.~Talwar, ``{Connection management xAPP for O-RAN RIC: A graph neural network and reinforcement learning approach},'' in \emph{IEEE International Conference on Machine Learning and Applications}, 2021, pp. 936--941.

\bibitem{qazzaz2024machine}
M.~M. Qazzaz, {\L}.~Ku{\l}acz, A.~Kliks, S.~A. Zaidi, M.~Dryjanski, and D.~McLernon, ``Machine learning-based xapp for dynamic resource allocation in o-ran networks,'' \emph{arXiv preprint arXiv:2401.07643}, 2024.

\bibitem{LLM_logic}
M.~Cosler, C.~Hahn, D.~Mendoza, F.~Schmitt, and C.~Trippel, ``{nl2spec: Interactively Translating Unstructured Natural Language to Temporal Logics with Large Language Models},'' in \emph{International Conference on Computer Aided Verification}, 2023, p. 383–396.

\bibitem{lewis2020retrieval}
P.~Lewis, E.~Perez, A.~Piktus, F.~Petroni, V.~Karpukhin, N.~Goyal, H.~K{\"u}ttler, M.~Lewis, W.-t. Yih, T.~Rockt{\"a}schel \emph{et~al.}, ``Retrieval-augmented generation for knowledge-intensive nlp tasks,'' \emph{Advances in Neural Information Processing Systems}, vol.~33, pp. 9459--9474, 2020.

\bibitem{lin2023pushing}
Z.~Lin, G.~Qu, Q.~Chen, X.~Chen, Z.~Chen, and K.~Huang, ``{Pushing large language models to the 6g edge: Vision, challenges, and opportunities},'' \emph{arXiv preprint arXiv:2309.16739}, 2023.

\bibitem{LLM_6G}
M.~Xu, D.~Niyato, J.~Kang, Z.~Xiong, S.~Mao, Z.~Han, D.~Kim, and K.~Lataief, ``{When Large Language Model Agents Meet 6G Networks: Perception, Grounding, and Alignment},'' \emph{arXiv: 2401.07764}, 2024.

\bibitem{tang2023domain}
Y.~Tang, A.~A.~B. Da~Costa, X.~Zhang, I.~Patrick, S.~Khastgir, and P.~Jennings, ``Domain knowledge distillation from large language model: An empirical study in the autonomous driving domain,'' in \emph{IEEE International Conference on Intelligent Transportation Systems (ITSC)}, 2023, pp. 3893--3900.

\bibitem{charismiadis20233gpp}
A.-S. Charismiadis, J.~M. Salcines, D.~Tsolkas, D.~A. Guillen, and J.~G. Rodrigo, ``The 3gpp common api framework: Open-source release and application use cases,'' in \emph{2023 Joint European Conference on Networks and Communications \& 6G Summit (EuCNC/6G Summit)}.\hskip 1em plus 0.5em minus 0.4em\relax IEEE, 2023, pp. 472--477.

\bibitem{nvidia_oran_chatbot_multimodal}
\BIBentryALTinterwordspacing
Nvidia, ``{Multimodal O-RAN RAG Chatbot with NVIDIA AI Foundation Endpoints or NVIDIA NIM for LLMs},'' 2024, accessed on 14.08.2024. [Online]. Available: \url{https://github.com/NVIDIA/GenerativeAIExamples/tree/main/experimental/oran-chatbot-multimodal}
\BIBentrySTDinterwordspacing

\bibitem{bornea2024telco}
A.-L. Bornea, F.~Ayed, A.~De~Domenico, N.~Piovesan, and A.~Maatouk, ``Telco-rag: Navigating the challenges of retrieval-augmented language models for telecommunications,'' \emph{arXiv preprint arXiv:2404.15939}, 2024.

\bibitem{LLM_control}
B.~Rong and H.~Rutagemwa, ``{Leveraging Large Language Models for Intelligent Control of 6G Integrated TN-NTN with IoT Service},'' \emph{IEEE Network}, pp. 1--1, 2024.

\end{thebibliography}


\begin{IEEEbiographynophoto}{YUN TANG (Member, IEEE)}
is a post-doctoral research fellow with Cranfield University from March 2024. He obtained PhD in Jan 2023 at Nanyang Technological University, Singapore and joined the University of Warwick as a research fellow from Feb 2023 to Mar 2024. His works have been published at several top-tier international conferences, such as IEEE ITSC, IEEE ICRA, and ASE, and in journals such as IEEE TSE and IEEE TIV. Currently, his research focuses on distributed intelligence in future networks.
\end{IEEEbiographynophoto}

\begin{IEEEbiographynophoto}{WEISI GUO (Senior Member, IEEE)}
is a Professor of Human Machine Interface at Cranfield University and also with the Alan Turing Institute. Previously, he was with the University of Cambridge, the University of Sheffield, and the University of Warwick. He is the winner of several IEEE Best Paper and IET Innovation awards in communications and currently co-leads the EPSRC 6G Future Communications Hubs in Distributed Computing and EPSRC Trustworthy Autonomous Systems Node in Security. He has published over 150 journal papers, mostly on networking and autonomy, with papers in Nature, Nature Comm., and Nature Machine Intelligence. He serves as editor of 3 IEEE and 1 Royal Society journals and contributes to emerging standards in wireless and AI. 

\end{IEEEbiographynophoto}

\end{document}